\documentclass[lettersize,  journal]{IEEEtran}
\usepackage{amsmath,  amsfonts}
\usepackage{algorithmic}
\usepackage{latexsym}
\usepackage{amssymb}
\usepackage{algorithm}
\usepackage{array}
\usepackage[caption=false,  font=normalsize,  labelfont=sf,  textfont=sf]{subfig}
\usepackage{textcomp}
\usepackage{stfloats}
\usepackage{url}
\usepackage{verbatim}
\usepackage{graphicx}
\usepackage{cite}
\usepackage{amsmath}
\usepackage{float} 
\usepackage{hyperref} 
\begin{document}

\title{Conditional Diffusion Model-Driven Generative Channels \\ for Double RIS-Aided Wireless Systems}

\author{Yiyang Ni, Qi Zhang, Guangji Chen, Yan Cai, Jun Li,  Shi Jin   ~\IEEEmembership{}

\thanks{Yiyang Ni is with the Jiangsu Key Laboratory of Wireless Communications, Nanjing University of Posts and Telecommunications, Nanjing 210003, China, also with the Jiangsu Second Normal University and Jiangsu
Institute of Educational Science Research, Nanjing 210013, China (email:
niyy@jssnu.edu.cn).}
\thanks{Jun Li and Shi Jin are with the National Mobile Communications Research Laboratory, Southeast University, Nanjing 210096, China (e-mail: jun.li@seu.edu.cn, jinshi@seu.edu.cn).(\textit{Corresponding author:Jun Li})}
\thanks{Qi Zhang and Yan Cai are with the School of Communications and Information Engineering, Nanjing University of Posts and Telecommunications, Nanjing 210003, China (email:1224014636@njupt.edu.cn, caiy@njupt.edu.cn)}
\thanks{Guangji Chen is with the School of Electronic and Optical Engineering, Nanjing University of Science and Technology, Nanjing 210094, China (email:guangjichen@njust.edu.cn)}}

\maketitle
\begin{abstract}
With the development of  the upcoming sixth-generation networks (6G), reconfigurable intelligent surfaces (RISs)  have gained significant attention due to its ability of reconfiguring wireless channels via smart reflections. However, traditional channel state information (CSI)  acquisition techniques for double-RIS systems face challenges (e.g., high pilot overhead or multipath interference). This paper proposes a new channel generation method in double-RIS communication systems based on the tool of conditional diffusion model (CDM). The CDM is trained on synthetic channel data to capture channel characteristics. It addresses the limitations of traditional  CSI generation methods, such as insufficient model understanding capability and poor environmental adaptability. We provide a detailed analysis of the diffusion process for channel generation, and it is validated through simulations. The simulation results demonstrate that the proposed CDM based method outperforms traditional channel acquisition methods in terms of normalized mean squared error (NMSE). This method offers a new paradigm for channel acquisition in double-RIS systems, which is expected to improve the quality of channel acquisition with low pilot overhead.   
\end{abstract}


\section{Introduction}
\IEEEPARstart{W}{ITH} the rapid development of fifth-generation (5G) and the imminent arrival of sixth-generation (6G) wireless communication networks,  the demand for innovative technologies that enhance performance,  extend coverage,  and improve energy efficiency has become increasingly critical [1][2].  
Among these emerging solutions, reconfigurable intelligent surfaces (RISs) have gained prominence as a transformative technology that enables intelligent control of the radio propagation environment by dynamically adjusting the phase shifts of the incident signals. Composed of a large array of passive reflecting elements with ultra-low power consumption, RIS offers a new degree of freedom for wireless channel optimization and enhances network adaptability. 

Although significant progress has been made in single-RIS systems,  the performance  is often constrained by the size, placement, and limited adaptability of the RIS in dynamic environments. To address these limitations,  double-RIS has been proposed,  wherein two RISs are deployed cooperatively to jointly optimize the transceiver design and passive beamforming of RISs.  By strategically positioning and controlling both surfaces,  double-RIS systems can significantly enhance system flexibility and performance, leading to the improved  signal-to-noise ratio (SNR) and channel capacity [3].  This architecture has  attracted growing attention for its potential to improve spatial diversity,  exploit multipath propagation,  and extend coverage in next-generation wireless networks.  

However, unlocking the potential benefits of a high passive beamfomring gain provided by the double-RIS architecture relies on the accurate estimation of channel state information (CSI).  Precise CSI is crucial for optimizing RIS configurations,  enhancing  transmission efficiency,  and improving overall system throughput [4].  However,  traditional CSI estimation techniques,  including pilot-based and model-based methods,  often rely on oversimplified assumptions regarding channel characteristics and system dynamics,  which may fail in highly dynamic or complex propagation environments [5].  Moreover,  compared with single-RIS architecture,  presence of two RISs introduces a higher dimension of channel coefficients and more complex structures of channels.  These interactions significantly increase modeling complexity, further limiting the practicality and effectiveness of conventional estimation techniques in real-world scenarios.

Recently,  deep learning (DL) has emerged as a promising approach to address the limitations of conventional CSI estimation methods [6].  For example,  a convolutional neural network (CNN) was employed in [7] to learn the mapping between received signals and channel responses.  Similarly,  a conditional generative adversarial network (CGAN)-based method was proposed in [8] to enhance CSI estimation accuracy in RIS-assisted mmWave MIMO systems.
However,  despite their effectiveness, these approaches  suffer from two fundamental limitations: they fail to fully exploit the phase coherence properties of RIS elements.  

To overcome the aforementioned issues, we propose a novel channel acquisition framework for double‐RIS communication systems based on a conditional diffusion model (CDM) that reconstructs complete CSI from partial observations. By leveraging spatial correlations between channel parameters and conditioning on received pilot signals from only a subset of reflecting elements, our generative model can produce high‐fidelity channel realizations that faithfully capture multipath fading, interference patterns, and spatial correlations under diverse RIS configurations. Unlike conventional schemes that require exhaustive, element‐wise estimation, this CDM‐based approach directly leverages partial channel information, thereby significantly reducing pilot overhead while maintaining estimation accuracy. Simulation results demonstrate that our method consistently outperforms traditional deep learning techniques in normalized mean squared error (NMSE), highlighting its promise for system design, performance optimization, and real‐time operation in complex double‐RIS deployments where efficient and accurate CSI acquisition is essential.

\begin{figure}[htbp]
    \centering
    \includegraphics[
        trim=30 230 10 220,   
        clip,              
        width=0.40\textwidth
    ]{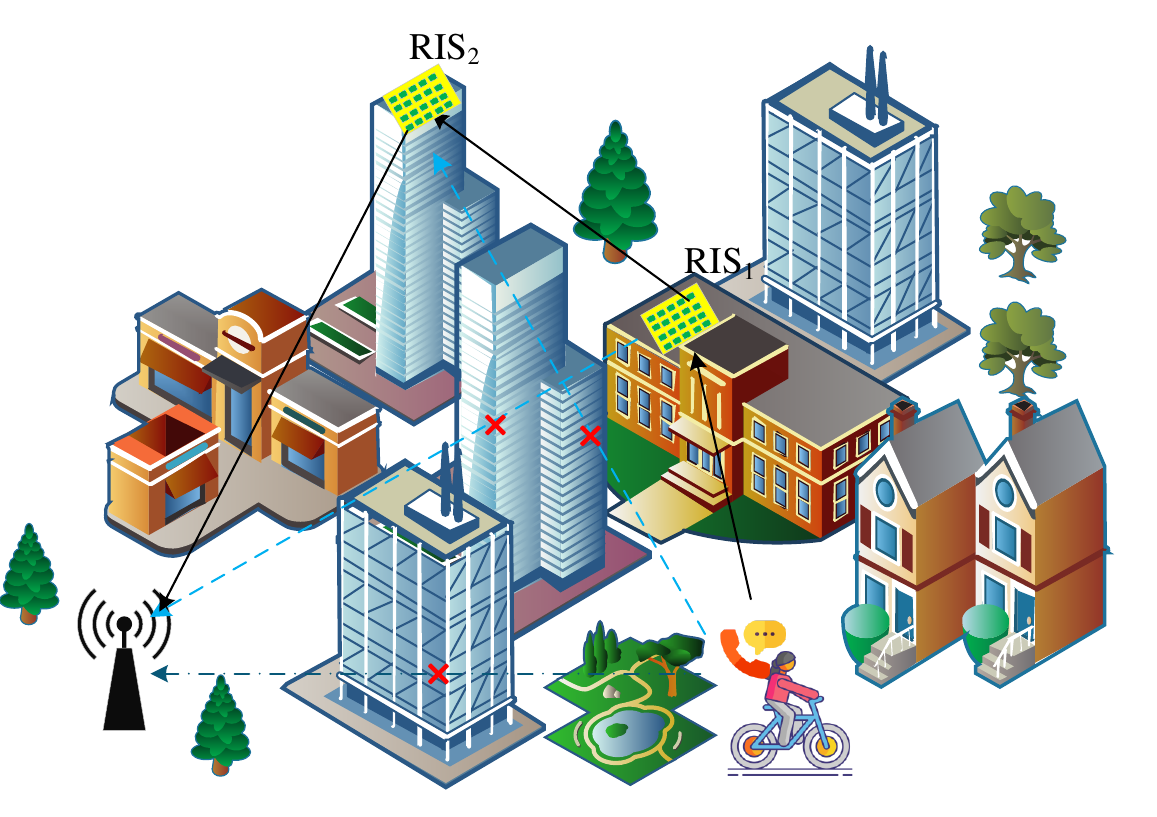} 
    \caption{A double RIS aided uplink wireless communication system.}
    \label{fig:1}
\end{figure}
\section{ Channel Model}
We consider a double-RIS-assisted communication system,  as illustrated in Fig.  1.  In this system,  a single-antenna user communicates with a base station (BS) equipped with $N$ antennas via two distributed RISs,  denoted as RIS$_1$ and RIS$_2$, respectively.  Due to the complexity of estimating the double-reflection link,  we focus solely on this path, assuming that the direct user-BS link is blocked by environmental obstructions (e.g., walls or corners in indoor scenarios).  To mitigate path loss and blockage effects,  RIS$_1$ is deployed near the user,  and RIS$_2$ is deployed close to the BS, enabling efficient communication via the double-reflection link.  Let $M_1$ and $M_2$ denote the number of passive elements at RIS$_1$ and RIS$_2$, respectively.  All channels are assumed to be quasi-static and follow a flat fading model within each coherence interval. 

We define the channel from the user to RIS$_1$  as $\mathbf{u} \triangleq \left[u_1, \cdots, u_m, \cdots,u_{M_1}\right]^T \in \mathbb{C}^{M_1 \times 1}$ where $u_m\left( m=1, \ldots, M_{1}\right)$ denotes the channel coefficient from the user to the $m$-th RIS$_1$ element. The channel between RIS$_1$  and RIS$_2$ is denoted by $\mathbf{D} \triangleq [\mathbf{d}_1, \cdots,\mathbf{d}_{m},\cdots,\mathbf{d}_{M_1}] \in \mathbb{C}^{M_2 \times M_1}$ where each column $\mathbf{d}_m = [d_{1,m}, \cdots,d_{m',m}, \cdots, d_{M_2,m}]^T\in \mathbb{C}^{M_2 \times 1}$ represents the channel coefficients from the $m$-th RIS$_1$ element to all RIS$_2$ elements. Here, $d_{m',m}$ is the channel coefficient between the $m$-th RIS$_1$ element and the $m'$-th $\left(m'=1, \ldots, M_{2}\right)$  RIS$_2$ element.
 In addition, $\mathbf{G}_2 \in \mathbb{C}^{N \times M_2}$ denotes the  channel from RIS$_2$ to the BS.The reflection coefficient vector of RIS$_\mu$ $\left(\mu\in\{1, 2\}\right)$ is denoted as $\boldsymbol{\theta}_{\mu}\triangleq \left[\theta_{\mu,  1},  \ldots,  \theta_{\mu,  M_{\mu}}\right]^{T}=\left[\beta_{\mu,  1}e^{j\phi_{\mu,  1}},  \ldots,  \beta_{\mu,  M_{\mu}}e^{j\phi_{\mu,  M_{\mu}}}\right]^{T}\in \mathbb{C}^{M_{\mu}\times1}$ where $\beta_{\mu,  m}=1$ and $\phi_{\mu,  m}\in \left[0, 2\pi\right)$ denote the reflection amplitude and phase shift of the $m$-th element of RIS$_1$,  respectively.  We consider a challenging scenario, where both the direct link and single-reflection links are severely blocked due to dense obstacles. To this end, we concentrate our analysis on this specific propagation path.  Accordingly,  the equivalent end-to-end channel between the user and the BS can be expressed as
\begin{align}
\textbf{h} = \textbf{G}_2 \boldsymbol{\Phi}_2 \textbf{D} \boldsymbol{\Phi}_1 \textbf{u}\label{align:1}
\end{align}
where $\boldsymbol{\Phi}_\mu = \mathrm{diag}\left(\boldsymbol{\theta}_\mu\right)$ denotes the diagonal reflection matrix of RIS$_\mu$.  Since we consider the fully passive RISs without signal reception or transmission capabilities,    it is infeasible to acquire the CSI between the two RISs.  To address this, we define the cascaded user $\rightarrow$ RIS$_1$ $\rightarrow$ RIS$_2$   channel as $\tilde{\textbf{D}} \triangleq \begin{bmatrix} \tilde{\textbf{d}}_{1},\cdots,\tilde{\textbf{d}}_{m},\cdots, \tilde{\textbf{d}}_{M_1} \end{bmatrix} = \textbf{D} \mathrm{diag}(\textbf{u}) \in \mathbb{C}^{M_2 \times M_1}$ where $\tilde{\textbf{d}}_{m} = \textbf{d}_m {u}_{m}\in \mathbb{C}^{M_2 \times 1}$ represents the contribution of the $m$-th RIS$_1$ element weighted by its user-side channel coefficient. Then, the channel model in (\ref{align:1}) can be equivalently expressed as
\begin{align}
\textbf{h} &= \textbf{G}_2 \boldsymbol{\Phi}_2 \tilde{\textbf{D}} \boldsymbol{\theta}_1 \notag\\
&= \textbf{G}_2 \left[\mathrm{diag}\left( \tilde{\textbf{d}}_{1} \right) \boldsymbol{\theta}_2,   \dots,   \mathrm{diag}\left( \tilde{\textbf{d}}_{ M_1} \right) \boldsymbol{\theta}_2 \right] \boldsymbol{\theta}_1 \notag\\
&= \sum_{m=1}^{M_1} \underbrace{ \textbf{G}_2 \mathrm{diag}( \tilde{\textbf{d}}_{ m} ) \boldsymbol{\theta}_2 }_{\textbf{B}_{m}} \theta_{1,  m}
\end{align}
where $\textbf{B}_{m} \in \mathbb{C}^{N \times M_2}$ denotes the effective channel associated with the $m$-th element of RIS, incorporating the complete path from the user to the BS via both RISs.
\section{Proposed Channel Generation Method}
In large-scale double-RIS systems,  the need to estimate CSI for numerous elements introduces substantial pilot overhead, which shortens the time available for data transmission and lowers system throughput. To mitigate this issue,  we estimate the CSI of a selected subset of elements, significantly reducing pilot overhead. The full CSI is then inferred by exploiting spatial correlations among channels.

In this section,  we propose a channel generation approach based on a CDM,  which consists of two key stages including the forward  and the reverse processes. The conditional diffusion model offers a robust framework for reconstructing complete CSI from partial channel state observations. 

We formulate the double-RIS channel generation  as a sampling process from  a learned latent prior,   where the double-RIS channel is reconstructed iteratively [9].  This process consists of main stages including a  forward process and a reverse process [10]. In the forward process,  double-RIS channel acquisition is modeled as a sampling procedurethat gradually transforms the initial data into a distribution resembling Gaussian noise.  The reverse process then iteratively denoises this data to reconstruct the complete channel. A conditional diffusion model is employed to generate channel realizations that closely match the actual distribution.  Let $\textbf{B}_{m}^{\text{P}} \in \mathbb{C}^{N\times M_{2}^{\text{P}}}$ denote the partial estimated cascaded channel corresponding to a subset of $M_2^{\text{P}}$ elements. The mask ratio is defined as $ \rho = {1-M_2^{\text{P}}}/{M_2}$,  indicating the fraction of elements with unestimated CSI.  Tuning the mask ratio enables a flexible balance between pilot overhead and channel acquisition  accuracy.

The partial cascaded channel $\textbf{B}_{m}^{\text{P}}$ is vectorized into a real-valued vector $\mathbf{x}_{0} \in \mathbb{R}^{2NM_2^{\text{P}}\times 1}$by stacking its real and imaginary components [11].  In the forward process, Gaussian noise is gradually added to $\mathbf{x}_{0}$, resulting in $\mathbf{x}_{T}$ after $\textit{T}$ diffusion steps. This process is formally defined as
\begin{align}\mathbf{x}_{t} = \sqrt{1 - \beta_{t}}\mathbf{x}_{t-1}+\sqrt{\beta_{t}}\boldsymbol{\epsilon}_{t},  \label{align:xx}\end{align}
\begin{figure*}[htbp] 
    \centering
    \includegraphics[
        trim=20 370 140 60,   
        clip,                         
        width=\textwidth            
    ]{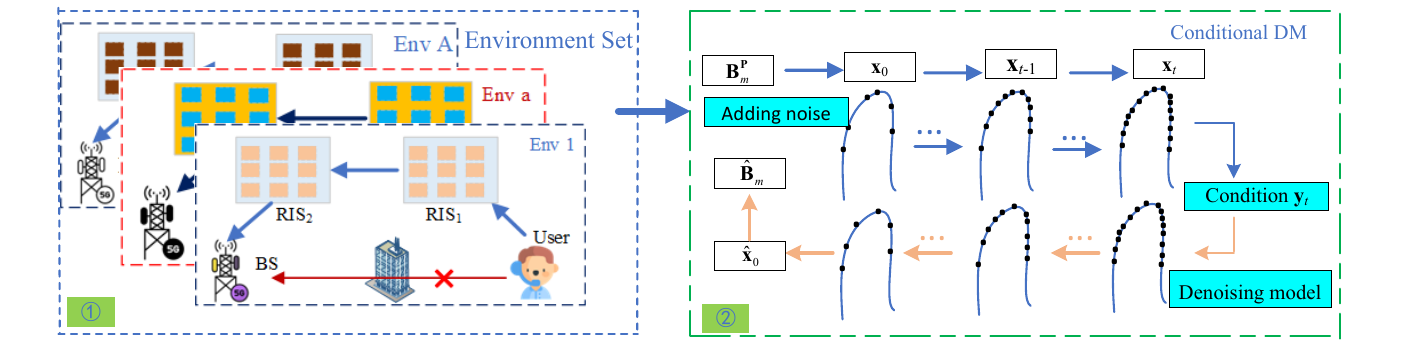} 
    \caption{Illustration of conditional diffusion model structure. It integrates a conditional diffusion model module,  which derives the complete channel state information from the partial channel state information by leveraging spatial correlations.  During the reverse process,  pilot signals are added as conditional input,  thereby significantly improving the generation effect.  }
    \label{fig:your_label}
\end{figure*}where $t \in\left[0,T\right]$ denotes the diffusion step, and the variance schedule $\beta_t$ increases linearly from $\beta_1$ to $\beta_T$,   and \(\boldsymbol{\epsilon}_{t}\) represents standard Gaussian noise. Based on the Markov chain framework,    the distribution of $\mathbf{x}_t$ conditioned on the original input $\mathbf{x}_0$ is given by
\begin{equation}p\left (\mathbf{x}_{t} \mid \mathbf{x}_{0}\right) =\mathcal{C N} \left(\sqrt{\bar{\alpha}}_{t} \mathbf{x}_{0},  \left(1-\bar{\alpha}_{t}\right)  \mathbf{I}\right) ,\label{align:4}\end{equation}
where $\overline{\alpha}_t = \prod_{i = 1}^{t} \alpha_i$,  $\alpha_t = 1 - \beta_t$ ,with $\mathbf{I}$ represents the identity matrix.  It is essential  for the subsequent training process to obtain $\textbf{x}_{t}$ by sampling from the distribution in   ($\ref{align:4}$) .  

During the conditional reverse process,  the reverse transition $p \left(\mathbf{x}_{t - 1}|\mathbf{x}_{t}\right) $ also follows Gaussian distribution,  and  can be written as
\begin{equation}p \left(\mathbf{x}_{t - 1}|\mathbf{x}_{t}\right)  = \mathcal{N} \left(\boldsymbol{\mu} \left(\mathbf{x}_{t},  t\right) ,  \boldsymbol{\Sigma} \left(\mathbf{x}_{t},  t\right) \right) ,  \end{equation}
where the mean and covariance are determined by the current state $\mathbf{x}_{t}$ and the time step $t$.  If the reverse distribution  $p \left(\mathbf{x}_{t-1}|\mathbf{x}_{t}\right) $ can be obtained during the denoising process,  the original data can be progressively reconstructed. 
To approximate this intractable distribution, we employ a neural network denoted as $p_{\theta} \left(\mathbf{x}_{t-1}|\mathbf{x}_{t}\right) $.  
  The corresponding mean and variance are respectively expressed as 
\begin{align}
{\mu}_{\theta} \left(\mathbf{x}_{t},  t\right) &=\frac{1}{\sqrt{{\alpha}_{t}}}\left(\mathbf{x}_{t}-\frac{\beta_{t}}{\sqrt{1 - \bar{\alpha}_{t}}}\boldsymbol{\epsilon}_{\theta} \left(\mathbf{x}_{t},  t\right) \right) ,  \notag\\
{\Sigma}_{\theta} \left(\mathbf{x}_{t},  t\right) &=\frac{1 - \bar{\alpha}_{t - 1}}{1 - \bar{\alpha}_{t}}\beta_{t}, \label{align:6} 
\end{align}
where $\boldsymbol{\epsilon}_{\theta} \left(\mathbf{x}_{t},  t\right) $ denotes the predicted noise generated by the neural network, given the input $\mathbf{x}_{t}$ at time step $t$. 

\begin{algorithm}[H] 
\caption{Training Process of the Proposed Method}
\label{alg:dt_training}

\textbf{Input:} $\textbf{B}_{m}^{\text{P}}$,   $T$,   $\{\alpha_{t}\}$,   $\{\beta_{t}\}$,   and training epochs $K$, weighting coefficient $\lambda_2$.

\quad $\textbf{B}_{m}^{\text{P}}$ collected from environments 1 to environments A.

\quad Initialization: Vectorize  $\textbf{B}_{m}^{\text{P}}$ into a real-valued vector $\textbf{x}_{0}$.

\quad Initialize U-Net parameters $\theta$ randomly.

\quad \textbf{for} $i = 1:K$ {do}

\quad \quad \textbf{for} $ t= 1:T$ {do}

$\quad$ $\quad$ Generate noise-corrupted $\mathbf{x}_{t}$ using $\{\alpha_t\}, \{\beta_t\}$ and $\boldsymbol{\epsilon}_{t}$ based on  ($\ref{align:xx}$). 

$\quad$ $\quad$ Extract $\textbf{y}_{t}$ for conditioning training data.

$\quad$ $\quad$ Train the U-Net to get $\tilde{\epsilon}_{\theta,t}$ based on  ($\ref{align:elta}$).

$\quad$ $\quad$ Compute loss function ${L}_{\theta}$ based on ($\ref{align:16}$).

$\quad$ $\quad$ Update $\theta$ using gradient descent on ${L}_{\theta}$.

$\quad$ $\quad$ \textbf{end for}

$\quad$ \textbf{end for}

\textbf{Output:} Trained noise prediction network parameters $\theta$.

\end{algorithm}
\begin{algorithm}[H] 
\caption{Inference Process of the Proposed Method}
\label{alg:dt_training}
\textbf{Input:} $\textbf{y}_{t}$,   $T$,   $\{\alpha_{t}\}$,   $\{\beta_{t}\}$,  weighting coefficient $\lambda_2$ and trained $\theta$.

$\quad$ Initialization: $\mathbf{x}_{t} \sim \mathcal{N} \left(\mathbf{0},  \mathbf{I}\right) $. 

$\quad$ \textbf{for} $t = T:1$ \textbf{do}

$\quad$ $\quad$ Update $\tilde{\epsilon}_{\theta,t}$.

$\quad$ $\quad$ Compute $\mu_\theta(\mathbf{x}_t, t)$  and $\Sigma_\theta(\mathbf{x}_t, t)$ based on  ($\ref{align:6}$). 

$\quad$ $\quad$ Compute $\mathbf{x}_{t-1}$ based on ($\ref{align:18}$).

$\quad$ \textbf{end for}

$\quad$ Convert real-valued vector $\mathbf{\hat{x}}_{0}$ to matrix $\hat{\mathbf{B}}_{m}$.

\textbf{Output:} Estimated ${\mathbf{\hat{B}}_{m}}$.
\end{algorithm}
Our objective is to train the model to accurately predict channel realizations based on the collected dataset.  As shown in [10], the training loss can be simplified to the Kullback–Leibler (KL) divergence between the true reverse distribution $p \left(\mathbf{x}_{t-1}|\mathbf{x}_{t}\right) $ and the approximation $p_{\theta} \left(\mathbf{x}_{t-1}|\mathbf{x}_{t}\right) $ at each diffusion step 
\begin{align}
L = \mathcal{D}_{\mathrm{KL}}\left[ p \left(\mathbf{x}_{t-1}|\mathbf{x}_{t}\right) || p_{\theta} \left(\mathbf{x}_{t-1}|\mathbf{x}_{t}\right) \right].
\end{align}

Conclusively,the training process is to minimize the discrepancy between the the noise predicted by the network and the true noise added during the forward process  
\begin{align}
L = \mathbb{E}\left[\left\|\boldsymbol{\epsilon}-\boldsymbol{\epsilon}_{\theta} \left(\mathbf{x}_{t},  t\right) \right\|^{2}\right].  
\end{align}

To generate high-quality channel samples,   relying solely on Gaussian noise to reconstruct $\mathbf{x}_0$ may yield random and inaccurate results. Therefore, we incorporate the received signal $\textbf{y}_{t}$, obtained from pilot transmissions, as an auxiliary input to the neural network.

In the double-RIS system,  the user transmits a pilot symbol $s$.   For simplicity, which is typically set to 1 for simplicity.   After propagation through the cascaded double-reflection links,  the received signal at the BS is given by
\begin{align}\textbf{y}_t = \textbf{G}_2\boldsymbol{\Phi}_2\textbf{D}\boldsymbol{\Phi}_1\textbf{u}s + \textbf{n},\end{align}
where $\textbf{n}\in\mathbb{C}^{N\times1}$ represents the additive noise.  The conditional reverse distribution is then redefined as $p \left(\mathbf{x}_{t-1}|\mathbf{x}_{t},  \mathbf{y}_{t}\right) $ which  can be expressed as follows by Bayesian theorem
\begin{align}
p\left(\mathbf{x}_{t - 1} | \mathbf{x}_{t},  \textbf{y}_{t}\right)&=\frac{p\left(\mathbf{x}_{t},  \textbf{y}_{t} | \mathbf{x}_{t - 1}\right)p\left(\mathbf{x}_{t - 1}\right)}{p\left(\mathbf{x}_{t},  \textbf{y}_{t}\right)}\notag\\\
&=\frac{p\left(\textbf{y}_{t} | \mathbf{x}_{t - 1},  \mathbf{x}_{t}\right)p\left(\mathbf{x}_{t} | \mathbf{x}_{t - 1}\right)p\left(\mathbf{x}_{t - 1}\right)}{p\left(\textbf{y}_{t} | \mathbf{x}_{t}\right)p\left(\mathbf{x}_{t}\right)}\notag\\\
&=\frac{p\left(\textbf{y}_{t} | \mathbf{x}_{t - 1},  \mathbf{x}_{t}\right)p\left(\mathbf{x}_{t - 1} | \mathbf{x}_{t}\right)}{p\left(\textbf{y}_{t} | \mathbf{x}_{t}\right)}\label{align:10}.
\end{align}
Since both $\mathbf{x}_{t}$ and ${\textbf{y}}_{t}$ are known during the denoising process, the term ${p \left({\textbf{y}}_{t}|\mathbf{x}_{t}\right) }$ is treated as a constant and denoted by $\lambda_1$.  Hence,   (\ref{align:10})  can be simplified as
\begin{align}p \left(\mathbf{x}_{t-1}|\mathbf{x}_{t},  {\textbf{y}}_{t}\right)  =  \lambda_{1}p \left({\textbf{y}}_{t}|\mathbf{x}_{t-1},  \mathbf{x}_{t}\right) p \left(\mathbf{x}_{t-1}|\mathbf{x}_{t}\right) ,  
\end{align}   
Furthermore, the likelihood term $p \left({\textbf{y}}_{t}|\mathbf{x}_{t-1},  \mathbf{x}_{t}\right) $ can else be expressed as
\begin{align}
p \left({\textbf{y}}_{t}|\mathbf{x}_{t-1},  \mathbf{x}_{t}\right) =\frac{p \left(\mathbf{x}_{t-1}|{\textbf{y}}_{t},  \mathbf{x}_{t}\right) p \left(\textbf{y}_{t}|\mathbf{x}_{t}\right) }{p \left(\mathbf{x}_{t}|\mathbf{x}_{t-1}\right) }.  \label{align:13}
\end{align}
Taking the logarithm of  (\ref{align:13}) and calculating the gradient,  we can obtain
\begin{align}
\nabla \log p \left(\textbf{y}_{t}|\mathbf{x}_{t - 1},  \mathbf{x}_{t}\right)\propto &\nabla \log p \left(\mathbf{x}_{t}|\textbf{y}_{t},  \mathbf{x}_{t - 1}\right) \notag\\
&- \nabla \log p \left(\mathbf{x}_{t - 1}|\mathbf{x}_{t}\right) .
\end{align}
We approximate the conditional distribution $p_{\theta} \left(\cdot\right) $ using a parameterized neural network ${p \left(\mathbf{x}_{t-1}|\mathbf{x}_{t},  \textbf{y}_{t}\right) }$, and thus obtain
\begin{align}
\nabla_{\theta} \log p_{\theta} \left(\mathbf{x}_{t-1}|\mathbf{x}_{t},  \textbf{y}_{t}\right)  \approx &\lambda_{2}\nabla_{\theta} \log p_{\theta} \left(\mathbf{x}_{t}|\textbf{y}_{t},  \mathbf{x}_{t-1}\right)  \notag\\
&+ \left(1 - \lambda_{2}\right) \nabla_{\theta} \log p_{\theta} \left(\mathbf{x}_{t-1}|\mathbf{x}_{t}\right) , 
\end{align}
where $\lambda_{2}$ is a  weighting coefficient that quantifies the importance of the conditional inputs $\textbf{y}_{t}$. Accordingly, we modify the loss function ${L}$ as
\begin{align}L \left(\theta\right) =\mathbb{E}\left[\left\|\boldsymbol{\epsilon}_{t} - \tilde{\boldsymbol{\epsilon}}_{\theta,  t}\right\|^{2}\right].  \label{align:16}
\end{align}
In (\ref{align:16}), $\tilde{\boldsymbol{\epsilon}}_{\theta,  t}$ is the modified predicted noise. 
\begin{align}
\tilde{\boldsymbol{\epsilon}}_{\theta,  t} = \lambda_{2}\boldsymbol{\epsilon}_{\theta} \left(\mathbf{x}_{t},  t, \textbf{y}_{t}\right) + \left(1 - \lambda_{2}\right) \boldsymbol{\epsilon}_{\theta} \left(\mathbf{x}_{t}, t\right) ,   \label{align:elta}
\end{align}
where $\boldsymbol{\epsilon}_{\theta} \left(\mathbf{x}_{t}, t,  \textbf{y}_{t}\right) $ represents the noise predicted with the additional conditional inputs $\textbf{y}_{t}$.  The update process from $\mathbf{x}_{t}$ to $\mathbf{x}_{t-1}$ can be represent as
\begin{align}
\mathbf{x}_{t - 1}=\frac{1}{\sqrt{\alpha_t}} \left(\mathbf{x}_{t}-\frac{\beta_{t}}{\sqrt{1 - \overline{\alpha}_{t}}}\tilde{\boldsymbol{\epsilon}}_{\theta,  t}\right) +\Sigma_{\theta} \left(\mathbf{x}_{t},t\right) z_{t}, \label{align:18}
\end{align}
where ${z}_{t}$ is a Gaussian noise.  We  encode the diffusion step ${T}$ and the received signal $\textbf{y}_{t}$ through the step and conditional embedding modules,  respectively.   These embeddings are then integrated with the noisy input  $\mathbf{x}_{t}$ and fed into a U-net.   A convolutional architecture is used to effectively fuse the three types of inputs.   Finally,   the network is trained by minimizing the loss function in (\ref{align:16}) .  The training and inference procedures are summarized in Algorithms 1 and 2,   respectively.  

\section{Simulation Results}
In this section,   we provide the simulation settings and results to evaluate effectiveness of the proposed method in double-RIS-assisted. The BS is equipped with a uniform linear array with $N = 4$ antennas.  To enable full CSI reconstruction from partial observations, we incorporate spatial correlation into the channel model [12]. Consider RIS$_1$ with $M_{1}$ reflecting elements, the spatial correlation matrix  $\Omega_{f, g}$  is defined as
\begin{align}\Omega_{f,  g}=\left\{
\begin{array}
{cc}1,   & \;k=0 \\
\frac{\sin  \left(k\right) }{k},   & \mathrm{otherwise},
\end{array}\right.  \end{align}
where $k=\frac{2\pi|f-g|d}{\lambda}$, $f,  g=0, 1, \cdots, M_{1}-1$, $d$ represents the distance between the  adjacent reflecting elements,  $\lambda$  represents the wavelength at which the system operates. Similarly,   the same correlation  model is applied to RIS$_2$. 

We now detail the network configuration for the proposed  CDM.  The diffusion model is trained with $\textit{T}$ = 500 steps,  and the noise $\beta_{t}$ is linearly increased from  $10^{-4}$ to 0.02.  During the reverse process, a U-Net based convolutional neural network is employed to approximate the denoising distribution at each step.  Furthermore, a conditional embedding module is utilized to incorporate the partial observed signals as auxiliary input.

NMSE is adopted as the metric of the estimation accuracy,   which is defined as
\begin{align}
\text{NMSE} = \mathbb{E}\left( \frac{\|\hat{\mathbf{B}}_{m} - \mathbf{B}_{m}\|_F^2}{\|\mathbf{B}_{m}\|_F^2} \right),
\end{align}
where $\hat{\mathbf{B}}_{m}$ and $\mathbf{B}_{m}$ represent the generative channel and the ground-truth,   respectively.  
\begin{figure}[htbp]
    \centering
    \includegraphics[width=1\linewidth, height=0.75\linewidth]{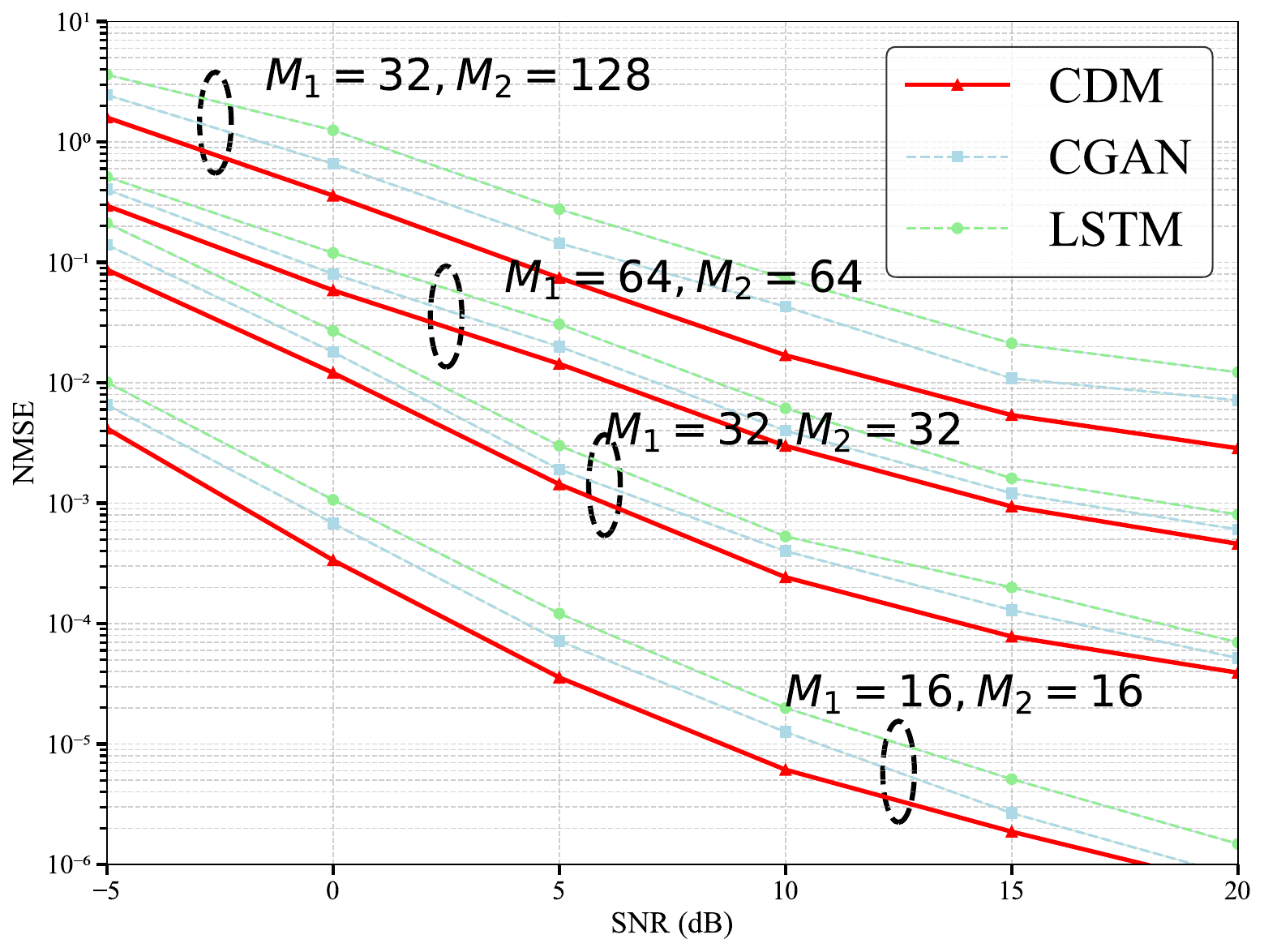}
    \caption{NMSE versus SNR under different $M_1$  and $M_2$. }
    \label{fig:3}
\end{figure}

As shown in Figure 3, we present the NMSE performance of the proposed CDM in the double-RIS system. For comparison, we also incorporate the long short-term memory (LSTM) estimation algorithm and a representative deep learning method, the CGAN. We evaluate the NMSE performance of the double-RIS system under different numbers of array elements. All evaluations are conducted within a SNR range from -5 dB to 20 dB, and comparisons with other methods are also made under different numbers of array elements. We set the values of mask ratio to 0.2 uniformly. It can be observed that, in terms of NMSE, the proposed CDM significantly outperforms both the LSTM estimation algorithm and the method based on the CGAN. This performance improvement is attributed to the ability of this method to reconstruct the complete CSI from partial CSI by exploiting spatial correlation, leveraging both the forward and reverse diffusion processes, while ensuring strict consistency between the training phase and the inference phase. As a result, the model gradually mitigates the impact of noise during the reconstruction process. Regarding the influence of the array size, we observe that the NMSE performance deteriorates as the number of array elements increases. This degradation of performance is caused by the increased complexity of the cascaded channel matrix between the two RISs, which substantially increases the computational burden.      

\begin{figure}[htbp]
    \centering

    \includegraphics[width=1\linewidth, height=0.75\linewidth]{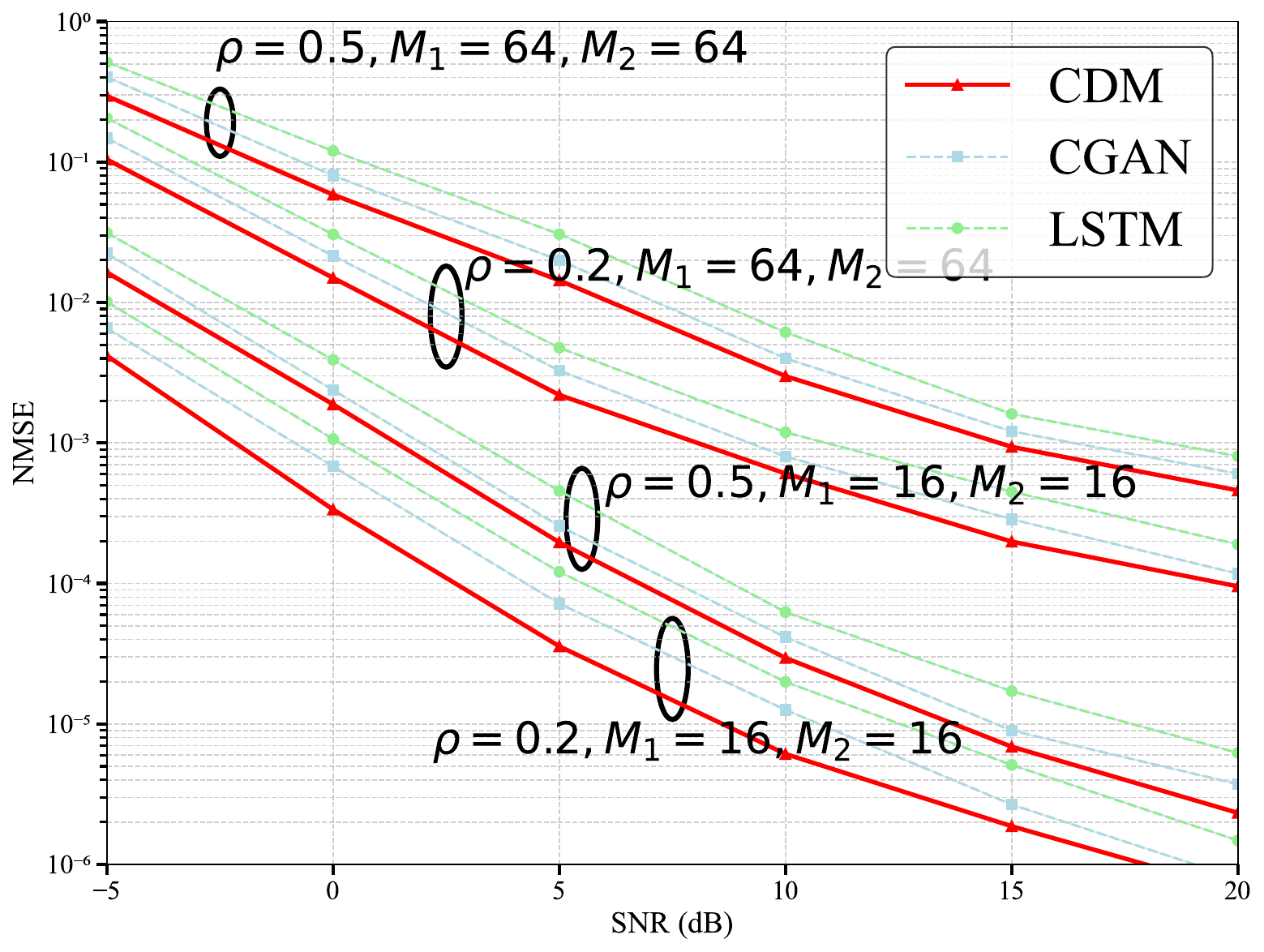} 
    \caption{NMSE versus SNR under different channel ratio $\rho$.}
    \label{fig:4}
\end{figure}

In Figure 4, we plot the comparison of the NMSE of the proposed CDM under different mask ratios and for different numbers of elements corresponding to these mask ratios. In this method, we use partial channel information for training to generate complete channel information. We compare the cases where the mask information ratios are 0.2 and 0.5 respectively, and also consider the cases where the number of array elements is 16 and 64 for these two ratios. It can be seen from the figure that as the SNR increases, the NMSE decreases, which is an expected result. In addition, at a fixed SNR level, the NMSE decreases as the partial information ratio increases. This phenomenon can be attributed to the fact that a higher ratio leads to an increase in the auxiliary information fed into the diffusion model network. Meanwhile, we also compare different methods under different mask  ratios. It can be seen from the figure that under the same mask ratio, the proposed CDM outperforms the CGAN and the  LSTM. This is because during the reverse denoising process of CDM, denoising is carried out step by step, enabling more complete utilization of the channel information, thus achieving better performance.      
\section{Conclusion}
In this paper, we proposed a novel double-RIS  channel generation method leveraging a conditional diffusion model. The CDM method utilizes spatial correlation to generate the full channel state information from partial channel state information and incorporates pilot signals as conditional inputs, resulting in a significant performance improvement. The proposed method abandons the traditional channel estimation methods that rely on prior theoretical models, and thus is more efficient in channel acquisition.

\end{document}